\documentclass[doublecol]{epl2}

\title{Detecting fractional Josephson effect through $4\pi$ phase slip}

\author{Zhen-Tao Zhang\inst{1}\thanks{zhzhentao@163.com} \and Zheng-Yuan Xue\inst{2} \and Yang Yu\inst{3}}
\shortauthor{Z. T. Zhang\etal}
\institute{
  \inst{1} Shandong Provincial Key Laboratory of Optical Communication, School of Physics Science and Information Engineering, Liaocheng University - Liaocheng 252059, PRC\\
  \inst{2} Guangdong Provincial Key Laboratory of Quantum Engineering and Quantum Materials, and School of Physics and Telecommunication Engineering, South China Normal University, Guangzhou 510006, PRC\\
  \inst{3} National Laboratory of Solid State Microstructures, School of Physics, Nanjing University, Nanjing 210093, PRC
  }

\abstract{Fractional Josephson effect is a unique character of Majorana Fermions in
topological superconductor system. This effect is very difficult to detect
experimentally because of the disturbance of quasiparticle poisoning and
unwanted couplings in the superconductor. Here, we propose a scheme to probe
fractional DC Josephson effect of semiconductor nanowire-based topological
Josephson junction through $4\pi $ phase slip. By exploiting a topological
RF SQUID system we find that the dominant contribution for Josephson
coupling comes from the interaction of Majorana Fermions, resulting the
resonant tunneling with $4\pi $ phase slip. Our calculations with
experimentally reachable parameters show that the time scale for detecting
the phase slip is two orders of magnitude shorter than the poisoning time of
nonequilibrium quasiparticles. Additionally, with a reasonable nanowire
length the $4\pi $ phase slip could overwhelm the topological trivial $2\pi $
phase slip. Our work is meaningful for exploring the effect of modest
quantum fluctuations of the phase of the superconductor on the topological
system, and provide a new method for quantum information processing.}

\begin{document}

\maketitle

\section{introduction}

Topological superconductor with $p$-wave pairing is a hot topic in condensed
matter physics. The system can host at its boundaries one kind of exotic
quasiparticles-Majorana Fermions (MFs), which are their own antiparticles.
MFs has important applications in quantum information processing \cite%
{Kitaev01,Nayak08,Xue13,Xue15}. Two separate MFs could construct one
physical qubit, named topological qubit. The non-locality makes topological
qubit immune from local environment noise. Nowadays, intrinsic topological
superconductor has yet to be found. Moreover, MFs are predicted to also
exist in some complicate systems, e.g., topological insulator coupled to
s-wave superconductor via proximity effect \cite{Fu08}, or spin-orbit
coupled semiconducting nanowire combined with superconductivity and magnetic
field \cite{Lutchyn10,Oreg10}. Recently, several groups have claimed that
they had observed some important signatures of MFs in these systems \cite%
{Mourik12,Deng12,Das12}. However, the existence of MFs has not been
confirmed due to the lack of a smoking-gun evidence.\newline
\indent A remarkable signature of MFs is fractional Josephson effect. It is
well known that the supercurrent through a conventional Josephson junction
is $2\pi $ periodic with the phase difference across the junction. However,
this statement is not always true for topological Josephson junction, which
is made with two weakly coupled topological superconductor instead of s-wave
superconductor. Kitaev has predicted that the current-phase relation in
topological Josephson junction should be $4\pi $ periodic \cite{Kitaev01}.
This period doubling of the Josephson current is protected by fermion parity
conservation. The fermion parity would not change unless a quasiparticle
excitation occurs. Unfortunately, non-equilibrium quasiparticles were found
in superconducting system at very low temperature, which is called
quasiparticle poisoning \cite{Matveev93,Joyez94}. It can break the parity
conservation of the system and restore the $2\pi $ period of the current in
the characteristic time. Therefore, the experiment to probe the $4\pi $
periodicity should be accomplished within the characteristic time of
quasiparticle poisoning. On the other side, the experimental duration time
is limited by adiabatic condition and measurement speed. Fast manipulation
of the phase difference can excite transitions from the subgap Majorana
bound states to the out-gap continuum states due to the Landau-Zener
transition. Therefore, it is challenge to experimentally detect the
fractional Josephson effect. Recently, several theoretical proposals are
brought forward to overcome the quasiparticle poisoning problem \cite%
{San-Jose12,Houzet13,Peng16}. Although these proposals are nearly insensitive to
quasiparticle poisoning, they all require that the junction works in the
ballistic regime, where the nanowire is nearly transparent, i.e., the
conductance $D\sim 1$. In this regime the nontopological Josephson junction
can also produce the fractional Josephson effect due to the Landau-Zener
transition \cite{Sau12,Sothmann13}. Therefore, it is desirable to figure out
a scheme working in the tunneling regime of the junction ($D\ll 1$). In
addition, most of previous researches have paid attentions to AC Josephson
effect where the junction is voltage or current biased. Actually, fractional DC 
Josephson effect, which does not bring dissipation, is more useful 
in the context of quantum information processing. For instance, it can be 
employed to couple topological qubits with conventional superconducting qubits. \newline
\indent Here we conceive a scheme for detecting fractional DC Josephson
effect. Compared with its AC analog \cite%
{Wiedenmann16,Bocquillon16,Bocquillon17}, the DC effect is more susceptible
to parity-breaking excitations and other imperfections. Generally, three
mechanisms, conventional Josephson coupling \cite{Pekker13}, quasiparticle
poisoning, the coupling of MFs from one topological superconductor, result a
conventional $2\pi $ phase slip which screens the $4\pi $ slip of
topological Josephson energy. By elaborately designing the parameters of
device and experiment, we can overcome these problems at the same time.
Firstly, the conventional Josephson coupling could be neglected when the
parameters of the superconducting circuit are proper, because the
conventional Josephson energy $E_{J}$ relies on the parameters of the
junction in a different manner with its topological analog $E_{m}$. When $%
E_{J}$ is much smaller than $E_{m}$, the $2\pi $ phase slips will be
inhibited. Secondly, our scheme can be implemented in a time scale much
shorter than the characteristic time of quasiparticle poisoning. At last,
the circuit used in our scheme could be reasonably designed such that the
interaction of MFs from one topological superconductor is much smaller than
the topological Josephson coupling. In this case, the $4\pi $ phase slip can
overwhelm the conventional $2\pi $ slip.
\section{System and Hamiltonian}

\begin{figure}[tbp]
 \centering
\includegraphics[scale=0.45]{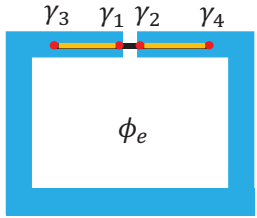}
\caption{(Color online) Schematic of the circuit. The superconducting loop is interrupted
by a superconductor-normal metal-superconductor junction. The loop is biased
with an external flux $\protect\phi _{e}$. The junction is formed by a
spin-orbit coupled nanowire laying on the separate superconductors. When the
magnetic field along the nanowire is larger than a critical value, the two
pieces of proximitized nanowire (orange sections) are topological
superconductors. At the boundaries are four MFs $\protect\gamma _{1},\protect%
\gamma _{2},\protect\gamma _{3},\protect\gamma _{4}$. }
\label{circuit}
\end{figure}

The system we considered is a superconducting loop interrupted by a
junction. The junction is made by putting a spin-orbit coupled semiconductor
nanowire on two separate superconductors. The two pieces of the nanowire
contacting with the superconductors underneath is superconducting due to
proximity effect. Combining with a parallel magnetic field, the nanowire
could be tuned into the topological phase. When the Zeeman splitting
excesses a critical value $B_{c}=\sqrt{\Delta ^{2}+\mu ^{2}}$ ($\Delta $ and
$\mu $ are$\ $the superconducting gap and the chemical potential,
respectively), the two pieces of proximitized nanowire will transition to
topological superconductors and two pairs of MFs emerge at their
boundaries(see Fig. \ref{circuit}). Moreover, the two MFs at the junction
couple with each other. The coupling Hamiltonian reads
\begin{equation}\label{eq1}
H_{m}=i\gamma _{1}\gamma _{2}E_{m}\cos \frac{\varphi }{2},
\end{equation}%
in which $\gamma _{1},\gamma _{2}$ are Majorana operators, $\varphi $ is the
phase difference across the junction, $E_{m}=\Delta \sqrt{D}$ is the
amplitude of the topological Josephson coupling energy with $D$ the
conductance of the quasi-one-dimensional nanowire. Besides, the conventional
Josephson coupling of the junction may also exist, which is related to the
quasi-continuum states above the superconducting gap. In the case of
one-channel nanowire, the conventional Josephson coupling can be written as
\begin{equation}\label{eq2}
H_{J}=-\Delta \sqrt{1-D\sin ^{2}\frac{\varphi }{2}}.
\end{equation}%
In the low conductance regime ($D\ll 1$), $H_{J}$ transforms to the
celebrated tunneling Josephson coupling $H_{J}=-E_{J}\cos \varphi $ (up to a
constant) with $E_{J}=\Delta D/4$. Therefore, it is straightforward to
deduce the relation $E_{J}=E_{m}^{2}/4\Delta $. If $E_{m}$ is much smaller
than the supercoducting gap, we can get $E_{J}\ll E_{m}$. In this case, we
can safely ignore $H_{J}$ term \cite{Hell16} and write the whole Hamiltonian
as
\begin{equation}\label{eq3}
H=E_{c}n^{2}+E_{L}(\varphi -\varphi _{e})^{2}+H_{m},
\end{equation}%
where $E_{c}=2e^{2}/C$ is the charge energy of the junction, and $%
E_{L}=(\phi _{0}/2\pi )^{2}/2L$ is inductive energy of the circuit with $%
\phi _{0}$ being flux quantum. $\varphi _{e}=2\pi \phi _{e}/\phi _{0}$, $%
\phi _{e}$ denotes the external flux threading the loop. The Hamiltonian is
as same as that of a flux qubit except the Josephson coupling term. As
well-known, a pair of MFs composes one Dirac fermion, and $H_{m}$ can be
expressed as
\begin{equation}\label{eq4}
H_{m}=E_{m}\cos \frac{\varphi }{2}(2f^{\dag }f-1),
\end{equation}%
in which we have defined $f=(\gamma _{1}+i\gamma _{2})/2$. The eigenvalue of
$f^{\dag }f$ (0 or 1) determines the parity of the Dirac fermion (even or
odd). The topological Josephson coupling given by $H_{m}$ has two
distinguishable characters. Firstly, the coupling is $4\pi $ periodic with
phase difference. Resultantly, the charge tunneling the junction is in unit
of single-electron instead of Cooper-pair. Very recently, an experiment \cite%
{Albrecht16} has examined this character in Coulomb blockade regime, in which $%
E_{C}\gg E_{m}$. In the opposite regime, that is $E_{C}\ll E_{m}$, the $4\pi
$ phase slip dual with single electron tunneling can occur. Secondly, the
coupling depends upon the fermion parity of the two MFs at the junction.
This character makes the $4\pi $ phase slip sensitive to the fermion-parity
breaking events, such as quasiparticle poisoning. In the following section,
we will present our scheme for uncovering the unique $4\pi $ feature of MFs.

\section{Scheme}

We now investigate how to observe the $4\pi $ phase slip with the system
shown in the last section. Without lossing generality, we assume that the
parity of MFs is restricted in the even subspace. Later on, we will consider
the effect of the unintended change of the parity on the phase slips. Under
the circumstances, the potential energy of the whole Hamiltonian (Eq. (\ref%
{eq3})) is
\begin{equation}\label{eq5}
U=E_{L}(\varphi -\varphi _{e})^{2}-E_{m}\cos \frac{\varphi }{2}.
\end{equation}%
By tuning the parameter $\varphi _{e}$ we can control the configuration of
the potential. If $\varphi _{e}=0$, the potential has one global minimum at $%
\varphi =0$ (see Fig.2A). If the flux is biased at $\varphi _{e}=2\pi $, a
symmetric double-well profile of the potential is formed, similar to the
potential of a flux qubit biased at $\varphi _{e}=\pi $. However, the
separation of the two minima of the double-well is $\sim $$4\pi $ instead of
$\sim $$2\pi $ (see Fig.2B). The lowest two energy eigenstates in the
double-well are symmetric and antisymmetric superpositions of left and right
local states. The energy splitting of them is denoted by $\Delta E$. For
probing the $4\pi $ phase slip, we initially set $\varphi _{e}=0$. In low
temperature limit, the system will be reset to the ground state in the well
around $\varphi =0$. Then, switch the bias to $\varphi _{e}=2\pi $ quickly
to make sure that the system localizes in the left well during this
operation, and wait for a time $\Delta t$ $\sim \frac{1}{\Delta E}$. In this
period, the resonant tunneling of the phase difference between the double
well can happen, and the state of the system is coherently oscillating
between the the left and right local state of the double well. Finally, bias
the circuit away from $\varphi _{e}=2\pi $ and measure the total flux of the
circuit. The resulting flux can either be about 0 or $2\phi _{0},$
corresponding to the left or right local state of the double well
respectively. The possibility of finding '$2\phi _{0}$' oscillates with $%
\Delta t$. In experiment, we can measure the total flux of the loop with
another RF SQUID \cite{Spanton17}. The possibility of the system projecting to
the $2\phi _{0}$ state can be obtained by repeating the above operations
many times. Note that if the same operations are applied to a conventional
or topological trivial RF SQUID, the final measured flux would definitely be
$\phi _{0}$ independent of $\Delta t$, because of the $2\pi $ periodicity of
their Josephson couplings \cite{Lange15}. Hence, the oscillating $4\pi $ phase
slip is a distinctive signature of topological Josephson junction. However,
in practice the superconducting circuit is subject to some unavoidable
disturbance which might destroy the signature. Therefore, it is vital to
investigate the robustness of our scheme.
\begin{figure}[tbp]
\includegraphics[width=7cm,height=4.5cm]{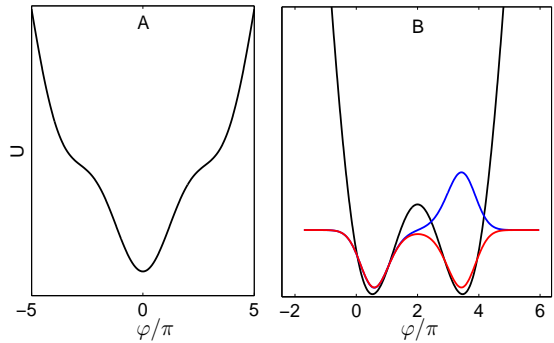}
\caption{Potential energy configurations. The fermion parity is even, and
the circuit is biased at $\protect\varphi _{e}=0$ (A), $\protect\varphi %
_{e}=2\protect\pi $ (B). In (B), the potential is a symmetric double-well
with the lowest two energy eigenstates be symmetric and antisymmetric
superpositions of left and right local states. }
\end{figure}

\subsection{Effect of quasiparticle poisoning}

In Eq. (\ref{eq5}), we have assumed that the parity of MFs is conserved in
the whole process. Actually, the parity conservation can be broken by
quasiparticle poisoning. Quasiparticles exist in various superconducting
systems even at vary low temperature. One quasiparticle excitation event
could alter the occupation of the in-gap states in a junction. For the
topological Josephson junction, it would turn over the parity of MFs. In our
case, we prepare the MFs at even parity state, thus an unwanted excitation
will take it to odd state. If this happens when the circuit is biased at $%
\varphi =2\pi $, the potential energy profile is changed. It is obvious that
the circuit will eventually stay at the ground state of the well with
minimum at $\varphi =2\pi $. That is exactly the result in conventional RF
SQUID in the same bias sequence. Thus, the $4\pi $ phase slip disappears.
Therefore, anyone who is going to observe the $4\pi $ phenomenon must carry
out the experiment in a period shorter than the quasiparticle poisoning
time. Generally, the parity lifetime of the bound state in a proximitized
semiconductor nanowire applied magnetic field exceeds 10 $\mu s$ \cite%
{Albrecht16}. The time needed to implement our scheme is on the order of $%
1/\Delta E$. Typically, we choose the parameters as follows: $E_{m}=25$ GHz$%
\times h$, $E_{c}=3$ GHz$\times h$, $E_{L}=1$ GHz$\times h$. With this
parameter configuration, we have numerically calculated the splitting $%
\Delta E=25$ MHz. This value means that the phase slips happen in the time
scale of $40$ $ns$, which is at least two orders of magnitude shorter than
the poisoning time. We stress that after each run of the experiment, the
Fermion parity will be initialized to even subspace. Therefore, we can claim
that the quasiparticles have little impact on our scheme.\newline
\indent A comment is in order. In our parameters set, the Josephson coupling
energy is much larger than the inductive energy with ratio $E_{m}/E_{L}=25$.
Even though, the finiteness of the ratio would make the distance of the two
minimum of the symmetric double well is not equal to $4\pi $, but rather
smaller than it. In fact, the distance is about $3\pi $ with our parameters.
From this view of point, the expression $4\pi $ phase slip is somewhat
misleading. Similarly, in a conventional RF SQUID the amplitude of the phase
slip is not $2\pi $ either ($<2\pi $). Actually, the names are stemming from
the formation of the relate Josephson coupling. What is more, we can
distinguish these two kinds of phase slips without any confusion.

\subsection{Effect of finite length of topological superconductor}

We know that the coupling of the two MFs of one topological superconductor
is oscillating with the length of the superconductor \cite{Cheng09,Sarma12}.
The oscillation amplitude decreases exponentially with the length $L$,
\begin{equation}\label{eq6}
\varepsilon =\varepsilon _{0}e^{-L/\xi },
\end{equation}%
where $\varepsilon _{0}$ is a prefactor, $\xi $ is superconducting coherence
length. Generally, if the topological superconductor is much longer than its
superconducting coherence length, this coupling is rather weak and can be
neglected. That is why we have not put the interaction between $\gamma _{1}$(%
$\gamma _{2}$) and $\gamma _{3}$($\gamma _{4}$) in Eq. (\ref{eq1}). However,
in practice, the length of a one-dimensional topological superconductor may
be limited by the technique to make it or the size of the circuit. It is
necessary to investigate the effect of the coupling between $\gamma _{1}$($%
\gamma _{2}$) and $\gamma _{3}$($\gamma _{4}$) on the $4\pi $ phase slips.%
\newline
\indent Let us first look at the Josephson coupling energy in absence of the
interactions $\gamma _{1}\gamma _{3}$, $\gamma _{2}\gamma _{4}$, ie., $H_{m}$
(Eq. (\ref{eq4})). When the phase difference takes values of $(2k+1)\pi $ (k
be integer), the even and odd parity states are degenerate. When the
interactions present, the potential energy can be addressed as
\begin{equation}\label{eq7}
U^{\prime }=E_{L}(\varphi -\varphi _{e})^{2}-E_{m}\cos \frac{\varphi }{2}%
\sigma _{z}+\varepsilon \sigma _{x},
\end{equation}%
\begin{figure}[tbp]
\includegraphics[width=7cm,height=6cm]{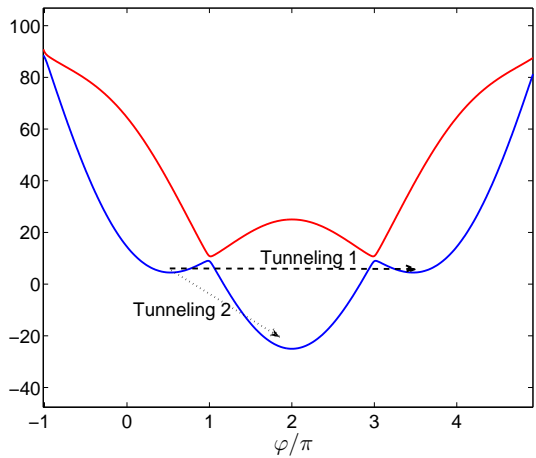}
\caption{Two kinds of tunnelings. The solid lines are describing potential
energy given by Eq. (\ref{eq7}) after diagonalization in the parity
subspace. The existence of MFs couplings $\protect\gamma _{1}\protect\gamma %
_{3},\protect\gamma _{2}\protect\gamma _{4}$ makes the transition of the
fermion parity of $\protect\gamma _{1}\protect\gamma _{2}$ possible.
Tunneling 1 (dashed line) does not change the parity while Tunneling 2
(doted line) does.}
\end{figure}
where $\sigma _{x,z}$ are Pauli operators acting in the fermion parity space
of $\gamma _{1},\gamma _{2}$. $\varepsilon $ denotes the coupling strength
of $\gamma _{1}\gamma _{3}$ ($\gamma _{2}\gamma _{4}$) which is much smaller
than $E_{m}$. It is easy to see that the odd-even degeneracies at $\varphi
=(2k+1)\pi $ are lift, and instead anticrossings arise, which leads to the
mixing of the two parity states. When the circuit is biased at $\varphi
_{e}=2\pi $ with the initial state be the ground state in the left well,
there are two possible tunneling events. One is tunneling to the right well
with same parity (named Tunneling 1), and the other is tunneling to the
nearest well with opposite parity (Tunneling 2), as shown in Fig. 3.
Tunneling 1 is the consequence of the topological Josephson coupling and
signify the $4\pi $ phase slip. In contrast, Tunneling 2 denote the $2\pi $
phase slip which is always connected to the topological trivial Josephson
junction. Therefore, if Tunneling 2 dominates the process, $4\pi $ phase
slip is covered and we can not tell the topological phase from the
topological trivial phase. To this end, one needs to clarify whether the
Tunneling 2 is weak enough to be neglected under experimentally feasible
condition. \newline
\indent Now we devote to estimate the tunneling rate of Tunneling 2. The
coexistence of parity switching and quantum fluctuation of the phase
difference make the task troublesome. We solve this problem in a
quasiclassical manner. As Tunneling 2 will change the fermion parity, it is
reasonable to believe that the tunneling rate should be related to the
transition rate of the parity states when $\varphi $ is considered as a
classical quantity. While biasing the circuit at $\varphi _{e}=2\pi $, the
system is initially located at left well with minimum of $\sim \pi /2$ (not
0 due to the finite of $E_{m}/E_{L}$) and parity is even. After Tunneling 2,
the system localizes at $\varphi =2\pi $ and parity is odd. Therefore, the
tunneling rate is limited by the transition rate of the fermion parity at $%
\varphi =\pi /2$. For convenience, we assume they are approximately equal.
The calculation of parity transition rate is a typical two-level-system
problem. Starting with even parity, the population of odd parity state is
oscillating with time between 0 and $P$, with $P=\varepsilon \Big /\sqrt{%
\varepsilon ^{2}+(E_{m}\cos \frac{\pi }{4})^{2}}$. According to Eq. (\ref%
{eq6}) and the parameters in Ref. \cite{Deng16}, when the nanowire is as
long as $L=2$ $\mu m$ which is reachable in experiment, the MFs coupling $%
\varepsilon $ is three orders of magnitude smaller than $E_{m}$. In this
case, the maximum odd parity population $P\approx 0$, which means that even$%
\rightarrow $odd transition rate is almost vanishing. One may argue that the
initial state does not localize at $\varphi =\pi /2$, but spreads on a range
even including the anticrossing point $\varphi =\pi $. In fact, the parity
transition rate reach its maximum value of $\varepsilon $ at the
anticrossing, which is the same order of magnitude as tunneling rate of
Tunneling 1, i.e., $\Delta E$. However, the probability of the initial state
be around the anticrossing is very small due to the large ratio $%
E_{m}/\varepsilon $, thereby Tunneling 2 would rarely occur in the period of
Tunneling 1. In other words, $4\pi $ phase slip will not be covered by $2\pi
$ phase slip.

\section{Discussion and Conclusion}

We would like to discuss the feasibility of our scheme. The scheme is
conceived based on the Hamiltonian of the system given by Eq. (\ref{eq3}),
in which we have neglected the conventional Josephson coupling of the
topological junction. For justifying this approximation, we estimate the
ratio $E_{J}/E_{m}$ with practical parameters. For the typical material NbN,
its superconducting critical temperature is $\sim $10 K, which equals eight
times of the value of $E_{m}$ chosen in this paper. This condition in turn
leads to $E_{J}=E_{m}/32$. Consequently, the conventional Josephson coupling
has little effect on the 4$\pi $ phase slips and can be ignored. In
addition, the large ratio of $\Delta /E_{m}$ is helpful to prevent the
subgap Majorana bound state being excited to the continuum states. The other
issue is the viability of RF SQUID with a very small inductance energy. It
is worth noting that a small value of the inductance energy and, thus, a
large magnitude of \emph{L} is essential for the observation of the $4\pi $
phase slip, since the large ratio $E_{m}/E_{L}$ can make the distance of the
minima of the double well of the superconducting phase far exceed $2\pi $.
Taken $E_{L}=1$ GHz, the inductance of the loop \emph{L} is up to 100 nH. In
experiment, we can design a large area superconducting loop, or make use of
a array of Josepshon junctions playing the role of a superinductor, such
as that in fluxonium qubit \cite{Pekker13}. Indeed, the requirement of the
large inductance could be loosed at the expense of slightly reducing the
amplitude of the phase slip.\newline
\indent In conclusion, we have proposed a scheme for detecting fractional DC
Josephson effect in topological RF SQUID system through $4\pi $ phase slip.
To observe this phase slip, we take advantage of the resonant tunneling of
the phase difference. Our calculations with reachable parameters show that
the duration of the process of the scheme is much shorter than the
quasiparticle poisoning time. More importantly, the $4\pi $ phase slip could
overwhelm the topological trivial $2\pi $ phase slip with a practical
nanowire length. Our scheme is experimentally feasible, and promising for
exploring the interplay of topological superconductors and quantum
computation.

\acknowledgments
We thank the very helpful discussions with Shi-Liang Zhu. This work was
funded by the National Science Foundation of China (No.11404156), the
Startup Foundation of Liaocheng University (Grant No.318051325), the NFRPC
(Grant No.2013CB921804), and the NKRDP of China (Grant No. 2016YFA0301800).

\end{document}